\documentclass[preprint,superscriptaddress,showpacs,preprintnumbers,amsmath,amssymb]{revtex4}

\usepackage{graphicx,color}
\usepackage{dcolumn}
\usepackage{bm}
\usepackage{multirow}

\begin{document}


\title{Influence of nuclear physics inputs and astrophysical conditions on \\Th/U chronometer}

\author{Zhongming Niu}
 \affiliation{State Key Laboratory of Nuclear Physics and Technology, School of Physics, \\ Peking University, Beijing 100871, China}

\author{Baohua Sun}
 \affiliation{School of Physics and Nuclear Energy Engineering, Beihang University, Beijing 100191, China}
 \affiliation{Justus-Liebig-Universit\"{a}t Giessen, Heinrich-Buff-Ring 14, Giessen 35392, Germany}

\author{Jie Meng}\thanks{e-mail: mengj@pku.edu.cn}
 \affiliation{School of Physics and Nuclear Energy Engineering, Beihang University, Beijing 100191, China}
 \affiliation{State Key Laboratory of Nuclear Physics and Technology, School of Physics, \\ Peking University, Beijing 100871, China}
 \affiliation{Department of Physics, University of Stellenbosch, Stellenbosch, South Africa}

\date{\today}

\begin{abstract}
The productions of thorium and uranium are key ingredients in
$r$-process nucleo-cosmochronology. With the combination of improved
nuclear and stellar data, we have made detailed investigations on
the $r$-process abundance pattern in the very metal-poor halo stars
based on the classical $r$-process approach. It is found that the
results are almost independent of specified simulations to observed
abundances. The influence from nuclear mass uncertainties on Th/U
chronometer can approach 2 Gyr. Moreover, the ages of the metal-poor
stars HE 1523-0901, CS 31082-001, and BD +17$^\circ$3248 are
determined as $11.8\pm 3.7$, $13.5\pm 2.9$, and $10.9 \pm 2.9$ Gyr,
respectively. The results can serve as an independent check for age
estimate of the universe.
\end{abstract}
\maketitle

\section{Introduction}

The age of the universe is one of the most important physical
quantities in cosmology. As the very metal-poor
([Fe/H]$\equiv\log_{10}(\mbox{Fe/H})_\ast-\log_{10}(\mbox{Fe/H})_\odot<$-2)
stars were formed at the early epoch of the universe, their ages can
set a lower limit on the age of the universe. The ages of these
stars can be determined by nuclear chronometers, which rely on the
comparison of the present abundances of radioactive nuclei with the
initial abundances at their productions. This method can avoid
uncertainties in the Galactic chemical evolution models, and thus
can be used as an independent dating technique for the universe.

The radioactive element Th was detected in the $r$-process enhanced
metal-poor star CS 22892-052 for the first time in
1995~\cite{CS22892052a}, and was observed later as well in other
metal-poor stars~\cite{Cowan1999ApJ,HD115444,HD221170}. These
observations make it possible to determine their ages using Th/X (X
represents a stable $r$-process element, e.g., Eu)
chronometer~\cite{Cowan1997ApJ}. The U line was firstly detected in
CS 31082-001~\cite{CS31082001}, which allows determining its age
using a more ideal Th/U~\cite{footnote} chronometer, since Th and U
are both $r$-only nuclei and close in mass number. Recently, the U
lines were also detected in other two metal-poor stars, namely, BD
+17$^\circ$3248~\cite{BD173248} and HE 1523-0901~\cite{HE15230901}.
These observations from metal-poor stars have provided the
opportunity of age estimates from Th/U chronometers. However, a
precise age determination via nuclear chronometers still requires
reliable estimations of the initial $r$-process abundances of
actinides.

The initial abundances of Th and U can be derived from $r$-process
nucleosynthesis calculations. However, the astrophysical condition
of $r$-process is still in debate up to now. It is found that the
$r$-process elemental abundance patterns observed from very
metal-poor stars and the solar system are universal for the heavier
elements above Ba. This universality suggests that there is probably
only one $r$-process site in the Galaxy
\cite{Cowan1999ApJ,Cowan2006Nature}, whereas a different process
could be responsible for the scatter of the lighter $r$-process
elemental ($38 \leqslant Z \leqslant 47$, here Z is atomic number)
abundance pattern \cite{Travaglio2004ApJ,Qian2007PRp,Montes2007ApJ}.
Furthermore, there are some indications that the observed stable
abundance pattern of the main r-process (except for U and Th)
extends to the light r-process elements
(e.g.,~\cite{Montes2007ApJ}). This uncertainty of astrophysical
site(s) has complicated $r$-process calculations, thus parameterized
$r$-process models
(e.g.,~\cite{Kratz1993ApJ,Pfeiffer1997ZPA,Arnould2007PR}) have been
widely employed to calculate the initial abundances of Th and U via
the best fit to the observed abundances.

A number of
investigations~\cite{Cowan1999ApJ,Goriely2001AA,Schatz2002ApJ,Kratz2007ApJ}
in $r$-process chronometers have been made so far in literatures.
These works employed the solar $r$-process abundances to predict the
zero-decay abundances of the radioactive elements (e.g., Th and U).
Nevertheless, a basic hypothesis of these works is that the
universal $r$-process abundance pattern from metal-poor stars and
the solar system not only hold for the elemental abundance
distribution ($56 \leqslant Z < 82$) but also for the isotopic
abundance distribution ($A \gtrsim 120$). However, with improved
observations of metal-poor stars
(e.g.,~\cite{CS31082001b,CS22892052,Sneden2008ARAA,
Sneden2009ApJSS}), now it is possible to make a step further for
$r$-process chronometers by directly simulating the abundances of
very metal-poor stars thus having a more solid ground. Furthermore,
by comparing the abundance patterns determined by simulating the
abundances of elements in the region $38 \leqslant Z \leqslant 82$
with those determined by simulating the abundances of elements in
the region $56 \leqslant Z \leqslant 82$, it may help to understand
the uncertainty of astrophysical site(s) of $r$-process
nucleosynthesis.

Another major source of the uncertainty of nuclear chronometer is
the nuclear physics inputs employed in $r$-process calculations,
such as nuclear masses and $\beta$-decay rates. Although lots of
efforts have been devoted in recent years to the measurements of
nuclear ground state properties, majority of neutron-rich nuclei of
relevance to the $r$-process are still out of the reach of
experimental capabilities. Consequently, theoretical predictions
have to be used. However, as there is no clear evidence to test the
reliability of theoretical models, it is necessary to make a
systematic investigation on the uncertainties of age estimates based
on different theoretical predictions, including those newly
developed models~\cite{Sun2008CPL}.

In this paper, we will analyze the abundance pattern in these
earliest Galactic stars and deduce their ages based on the
combination of improved stellar and nuclear data. The paper is
organized as follows. In Sect. II, numerical details including a
short introduction to the classical $r$-process model and the
nuclear physics inputs used in this paper are given. In Sect. III,
possible constraints on nuclear mass models are discussed based on
the experimental data and the solar $r$-process abundances. Then
calculations are made by directly simulating the abundances in
metal-poor stars. Moreover, the influence of astrophysical
conditions and nuclear physics inputs on the $r$-process simulations
and on the age estimates of the stars HE 1523-0901, CS 31082-001,
and BD +17$^\circ$3248 are discussed in details. Finally, the
summary and perspective are presented in Sect. IV.

\section{Sketch of the classical $R$-process model and numerical details}

The classical $r$-process model is employed in this investigation to
deduce the zero-decay Th/U abundance ratio similar to
Refs.~\cite{Cowan1999ApJ,Schatz2002ApJ}. In this model, seed-nuclei
(Fe) are irradiated by neutron sources of high and continuous
neutron densities $n_n$ over a timescale $\tau$ in a high
temperature environment ($T\sim$1 GK). Similar to
Refs.~\cite{Sun2008PRC, Pfeiffer1997ZPA}, sixteen components with
neutron densities in the range of $10^{20}$ to $3 \times 10^{27}$
cm$^{-3}$ are used to reproduce the observed $r$-process abundances.
The weight $\omega$ and the irradiation time $\tau$ of each
$r$-process component follow exponential relations on neutron
density $n_n$:
\begin{eqnarray}
\omega(n_n)=d\times n_n^a, \quad \tau(n_n)=b\times n_n^c \; .
\label{eq:relations}
\end{eqnarray}
The parameters $a$, $b$, and $c$ can be determined from a
least-square fit to the observed $r$-process abundances with the
Marquardt method~\cite{Marquardt1968JSIAM}, while $d$ follows the
normalization of the weighting factors. This model is considered as
a realistic simplification of dynamical $r$-process model, and it
has been successfully employed in describing $r$-process patterns of
both the solar system and metal-poor stars
(e.g.~\cite{Sun2008PRC,Kratz2007ApJ,Freiburghaus1999ApJ,Pfeiffer2001NPA}).

The abundances for each $r$-process component are calculated within
the waiting-point approximation. In this method, the abundance
distribution in an isotopic chain is given by the Saha equation and
is entirely determined by neutron separation energies for a given
temperature $T$ (in this paper, $T=1.5$ GK) and a neutron density
$n_n$. The matter flow between neighboring isotopic chains is
determined by the total $\beta$-decay rates (including up to three
neutrons $\beta$-delay emissions). After neutron sources freeze out,
all the isotopes then proceed to the corresponding stable isotopes
via $\beta$- and $\alpha$-decays. The process of spontaneous fission
is also taken into account to calculate the final abundances of
actinides.

During the early phase of the $r$-process, the waiting-point
approximation is generally believed to be well maintained
~\cite{Kratz1993ApJ, Freiburghaus1999ApJ} since the reaction rates
for neutron captures and photo-dissociations are much faster than
the $\beta$-decay rates. Recent dynamical calculations showed that
the neutron captures may influence the final abundance distribution
(e.g.~\cite{Surman2009PRC}) where the corresponding neutron capture
rates varied by 2 orders of magnitude. However, the Th/U ratio
should not be affected dramatically in comparison to the Th/X ratio
by the waiting-point approximation. Other effects induced by
neutrino spallation~\cite{Qian1997PRC} during the freeze out may
have small influence on the Th and U
chronometer~\cite{Terasawa2004ApJ}. Therefore, we will focus
hereafter on the influence of the nuclear ground state properties
within the waiting-point approximation.

The astrophysical condition of $r$-process nucleosythesis can be
obtained by the best fit to the observed abundances in metal-poor
stars or the solar $r$-process abundances. In this paper, we take
the scaled average abundances of the most Eu-enriched stars CS
31082-001~\cite{CS31082001b} and CS 22892-052~\cite{CS22892052} as a
representative of stable, universal $r$-process abundances for
elements in the region $38 \leqslant Z < 82$. However, there is an
obvious discrepancy of Pb abundance in these two metal-poor stars,
which may be due to the obscure $s$-process contribution
\cite{Eck2001Nature}. Hence in order to estimate the uncertainty of
Pb abundance and furthermore its influence on the actinide abundance
predictions, in our first step we have adopted the Pb abundances in
these two stars to find out their corresponding best simulations.
The corresponding results will be discussed in next section. As the
Pb abundance in CS 22892-052 is in better agreement with the scaled
solar $r$-process abundance, the Pb abundance hereafter is adopted
from CS 22892-052 otherwise specified.

In order to estimate the potential astrophysical uncertainties, we
have performed different fits to the observations. In the first
case, all the neutron-capture elements in the very metal-poor stars
are assumed to be synthesized in a unique $r$-process site.
Consequently, one has to fit the average scaled abundances of
elements in the range $38 \leqslant Z \leqslant 82$ (labeled as fit
I hereafter). In the second case, which states that the abundance
pattern of heavier $r$-process elements ($Z \geqslant 56$) in
metal-poor stars are identical to those in the solar system, we
perform a fit to the average scaled abundances of elements only in
the range $56 \leqslant Z \leqslant 82$ (labeled as fit II), as well
as a fit to the solar $r$-process isotopic abundances with mass
number in the region $125 \leqslant A \leqslant
209$~\cite{Sneden2008ARAA} (labeled as fit III). The results from
the fit I and II can help to understand whether there are different
origins for elements lighter and heavier than Barium. Meanwhile, the
two independent calculations (the fit II and III), in principle, may
serve as a critical test to the hypothesis, namely, whether the
isotopic abundances are identical in metal-poor stars and the solar
system.

In this paper, available experimental
data~\cite{Audi2003NPA:a,Sun2008NPA} are used, otherwise predictions
of three newly developed mass models (HFB-17~\cite{HFB17},
RMF~\cite{RMF}, and KTUY~\cite{KTUY}) or five commonly used mass
models DZ10~\cite{DZ10}, DZ28~\cite{DZ28}, ETFSI-2~\cite{ETFSI2},
ETFSI-Q~\cite{ETFSIQ}, FRDM~\cite{FRDM} are employed. These models
span from macroscopic-microscopic to self-consistent microscopic
models. As for the $\beta$-decay rates, the predictions of the
FRDM+QRPA method~\cite{FRDMQRPA} are employed throughout the paper
as a complementary to the experimental data~\cite{Audi2003NPA:a},
and those from the ETFSI+CQRPA method~\cite{ETFSICQRPA} are used for
comparison.

\section{Results and discussions}

In this section, we will focus on the prediction of Th/U production
ratio, the age determination of metal-poor stars as well as their
uncertainties related to astrophysical conditions and nuclear
physics inputs. We will start with the global comparison and
possible constraint on nuclear mass models.

\subsection{Comparison and Constraint on Mass Models}

\begin{figure}
  \includegraphics[width=12cm]{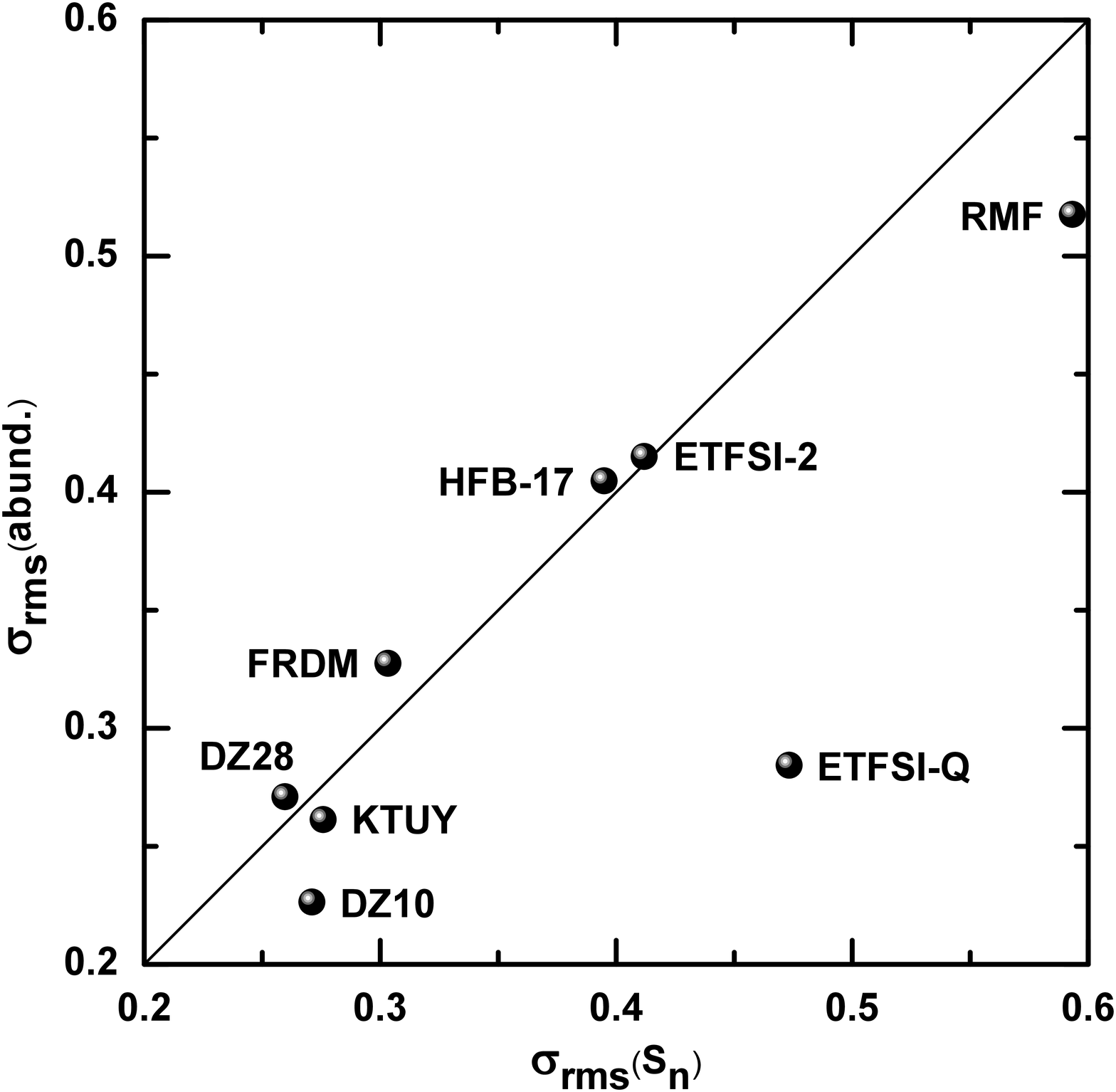}\\
  \caption{$\sigma_{\mbox{rms}}$(abund.) and $\sigma_{\mbox{rms}}(S_n)$ for
    different mass models. The lines are plotted to guide eyes.
    For details refer to the text.}\label{fig1}
\end{figure}

$R$-process calculations using various nuclear mass models may yield
abundances differing even by several orders of
magnitude~\cite{Sun2008PRC}. It is therefore essential to test their
reliability before applying them in the astrophysical calculations.
As one-neutron separation energy ($S_n$) is the practically used
quantity in $r$-process calculations, the rms deviations of $S_n$
($Z=26-100$) calculated in different mass models relative to
experimentally measured values~\cite{Audi2003NPA:a,Sun2008NPA},
$\sigma_{\mbox{rms}}(S_n)$, are calculated and shown by the
horizontal axis in Fig.~\ref{fig1}. In all models considered here,
$\sigma_{\mbox{rms}}(S_n)$ is within 0.6 MeV. The DZ28 model better
reproduces the experimental $S_n$, whereas the
$\sigma_{\mbox{rms}}(S_n)$ of the RMF model is as large as 0.59 MeV.
As for the latest version of the HFB model, a
$\sigma_{\mbox{rms}}(S_n)$ of 0.39 MeV is somewhat in between.
Furthermore, the best fits to the solar $r$-process abundances with
mass numbers in the range of 125 to 209 have been done using
different mass models (the fit III mentioned in Sec. II) and the
corresponding rms values relative to the solar $r$-process
abundances, $\sigma_{\mbox{rms}}$(abund.), are shown by the vertical
axis in Fig.~\ref{fig1}.

From Fig.~\ref{fig1}, it is clearly shown that the model, which
describes the experimental $S_n$ values better, generally results in
better agreement with the observations in the $r$-process
simulation. A clear deviation to this trend is observed for the
ETFSI-Q model. This mainly results from the phenomenological
quenching effect at the $N=126$ shell closure introduced in the
ETFSI-Q model, which smoothes out the trough before $A=195$ peak.
This effect is not predicted in any of the other
models~\cite{Sun2008PRC}. Moreover, it may be worth mentioning that
the $r$-process simulation is generally not improved using the
HFB-17 model, although its reproductive power of nuclear mass has
been significantly improved compared with the previous versions.
Potential problems in the HFB models, like the fitting procedure
employed to get the effective interaction parameters, have been
discussed in~\cite{Stone2005JPG}. Another open question in these
mean-field models is how to incorporate self-consistently the
effects beyond mean field. On the other hand, the results from DZ
models are surprisingly good, which may imply that such effects have
been effectively treated better in DZ models~\cite{Olofsson2006PRL}.

Constraints on various mass models, in principle, can also be
obtained from the present knowledge of the evolution of our solar
system. Using the present abundance ratio Th/U$=3.80$ from solar
system~\cite{Lodders2003ApJ} as well as their known half-lives
($\tau_{1/2}[\textrm{Th}]=1.405 \times 10^{10}$ yr and
$\tau_{1/2}[\textrm{U}]=4.468 \times 10^9$ yr~\cite{Audi2003NPA:a}),
the ratio can be calculated by extrapolation backward in time to an
epoch when solar system became a closed system, namely 4.6 Gyr
ago~\cite{Rolfs1988Book}. Beyond this time, the estimate of the
Galactic age has to rely on the Galactic chemical evolution model,
which is however not well understood yet. Instead, some constraints
can be obtained from two extreme assumptions of the Galactic
chemical evolution models: the sudden and uniform
synthesis~\cite{Rolfs1988Book}, where the former can provide a lower
boundary of the Galactic age.

\begin{figure}
  \includegraphics[width=12cm]{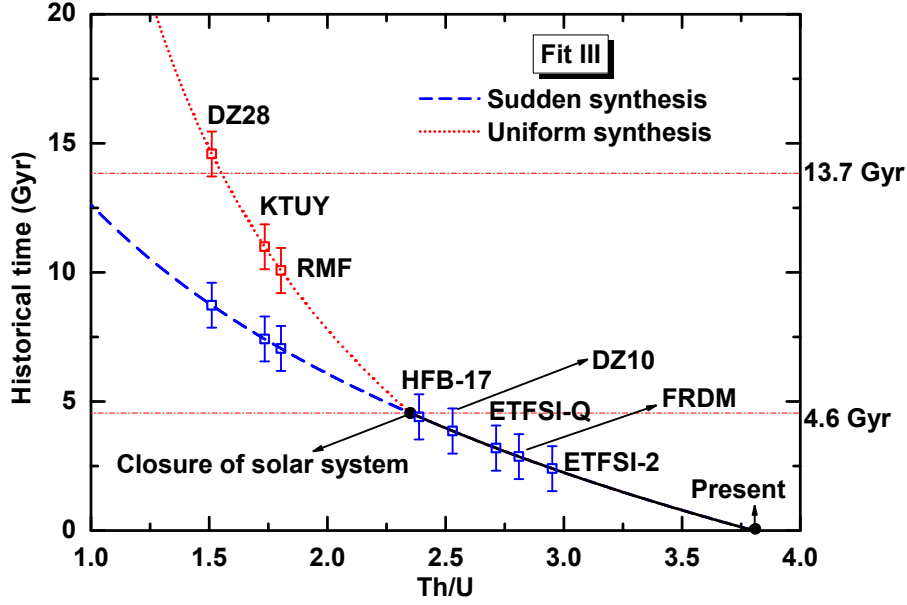}\\
  \caption{(Color online) The historical time and the corresponding abundance ratio Th/U.
    The solid line between the present and the time when the solar system became
    a closed system can be deduced from the radioactive decay of Th and U.
    The predictions from the assumptions of sudden and uniform synthesis
    are shown by the dashed and dotted curves, respectively. The predicted zero-age Th/U ratios
    using different mass models are calculated based on the fit III,
    and are marked in the corresponding curves. The dash-dotted lines represent the age of the
    solar system (4.6 Gyr~\cite{Rolfs1988Book}) and the age of the universe (13.7 Gyr~\cite{WMAP}).
    For details refer to the text.}\label{fig2}
\end{figure}

The time evolutions of the abundance ratio Th/U are governed by
assuming the sudden and uniform synthesis, respectively,
\begin{eqnarray}
  \left(\frac{\textrm{Th}}{\textrm{U}}\right)_{\textrm{SS}} &=& \left(\frac{\textrm{Th}}{\textrm{U}}\right)_{\textrm{r-process}}
       e^{-(\lambda_{\textrm{Th}}-\lambda_{\textrm{U}})(t-4.6)}~\quad \textrm{(sudden synthesis)},\label{eq:sudden}\\
  \left(\frac{\textrm{Th}}{\textrm{U}}\right)_{\textrm{SS}} &=& \left(\frac{\textrm{Th}}{\textrm{U}}\right)_{\textrm{r-process}}
       \frac{\lambda_{\textrm{U}}[1-e^{-\lambda_{\textrm{Th}}(t-4.6)}]}{\lambda_{\textrm{Th}}[1-e^{-\lambda_{\textrm{U}}(t-4.6)}]}~\quad \textrm{(uniform
       synthesis)},\label{eq:uniform}
\end{eqnarray}
where $\lambda=\ln 2/\tau_{1/2}$ is the corresponding decay
constant, $(\textrm{Th}/\textrm{U})_{\textrm{SS}},
(\textrm{Th}/\textrm{U})_{\textrm{r-process}}$ represent the
primordial solar system abundance ratio and the ratio after
r-process nucleosynthesis. From these two equations, the Th/U ratio
can be deduced as a function of time for both assumptions with
$(\textrm{Th}/\textrm{U})_{\textrm{SS}}=2.35$~\cite{Lodders2003ApJ},
as shown by the dashed and dotted curves in Fig.~\ref{fig2}. On the
other hand, the Th/U ratio after r-process nucleosynthesis can also
be estimated from simulating the solar r-process abundance pattern,
namely, the fit III. In this case, one can determine the
corresponding time for the epoch of nucleosynthesis using
Eqs.~(\ref{eq:sudden}) and (\ref{eq:uniform}). Different predictions
when using various mass models are labeled in the corresponding
curves. Clearly, a reasonable age estimate for the epoch of
nucleosynthesis should be longer than the age of the solar system
(4.6 Gyr) but shorter than the age of the universe (13.7
Gyr~\cite{WMAP}).

The simulations using the ETFSI-2, ETFSI-Q and FRDM models predict
too larger Th/U abundance ratio, thus the corresponding age
estimates are clearly lower than the age of the solar system even if
the observable uncertainty of 0.04 dex~\cite{Lodders2003ApJ} is
included. Especially for the frequently used ETFSI-Q model, the
present result, lower than the age of the solar system, is in
agreement with the age estimate in Ref.~\cite{Goriely2001AA} but in
disagreement with those in Refs.~\cite{Schatz2002ApJ,Kratz2007ApJ}.
On the other hand, the Th/U predictions from five other mass models,
DZ28, KTUY, RMF, HFB-17 and DZ10, are consistent with this
constraint of the Galactic age. Of course, a better knowledge of the
Galactic chemical evolution would further tight the constraint on
mass models.

\subsection{The Influence of Astrophysical Conditions and Mass Models}

\begin{figure}
  \includegraphics[width=15cm]{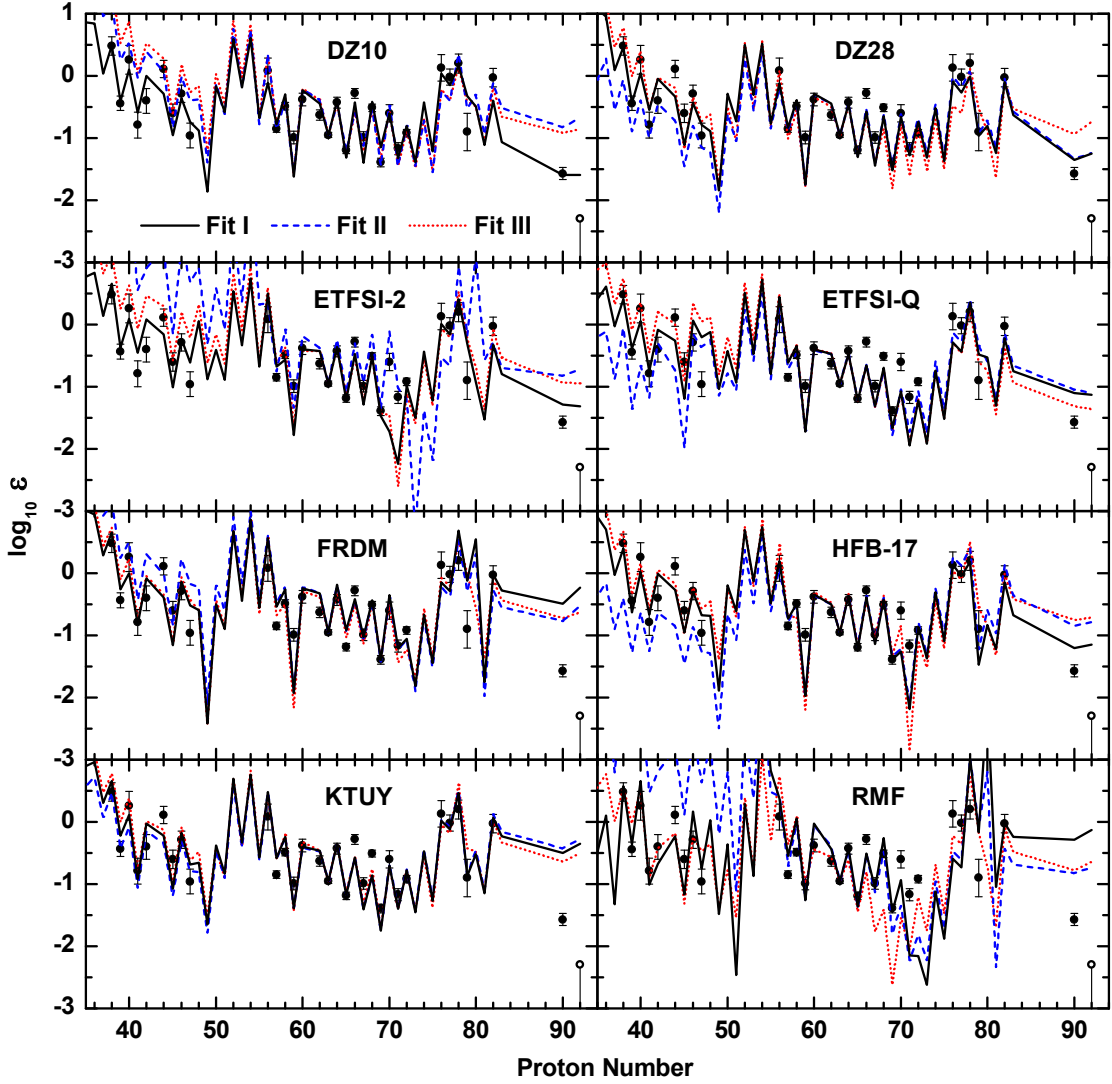}\\
  \caption{(Color online) Calculated $r$-process abundances (scaled to Eu)
    using various mass models. The $\beta$-decay properties are taken
    from the FRDM+QRPA model. The simulations from the fit I, II,
    and III are denoted by the solid, dashed, and dotted curves, respectively.
    The filled circles represent the scaled average element abundances of CS 31082-001 and CS
    22892-052 in the region $Z<82$. The abundances for Pb, Th and the upper limit (open circle) on
    U are taken from CS 22892-052. We adopt the usual notation
    that $\log_{10} \varepsilon(A) \equiv \log_{10}(N_A/N_H)+12.0$ for element $A$, where
    $N$ represents abundance.}\label{fig3}
\end{figure}

With the recent observations of metal-poor
stars~\cite{Sneden2008ARAA, Sneden2009ApJSS}, it is now possible to
determine the astrophysical condition by simulating directly the
abundances of $r$-process enhanced metal-poor stars. The simulations
from the fit I and II are shown for all the mass models in
Fig.~\ref{fig3} by the solid and dashed lines, respectively. For
comparison with previous investigations, simulations are also
preformed based on the solar $r$-process abundances, namely the fit
III (dotted lines).

In the region $56 \leqslant Z \leqslant 82$, where all the three
fits covered, all simulations well reproduce the stellar abundances.
For the light neutron-capture elements with $38 \leqslant Z
\leqslant 47$, the calculated abundances from the fit II are much
higher than those from the fit I when the RMF mass model is adopted.
Similar results have been also found for simulations using the DZ10,
ETFSI-2 and FRDM models. On the other hand, the calculated
abundances from the fit II are only slightly lower than those from
the fit I for the DZ28, ETFSI-Q, HFB-17 and KTUY simulations.
Therefore, it is hard to distinguish the "potential" difference
between the fit I and the fit II in the present work, since two
different conclusions can be drawn. As for the calculations from the
fit III, they are in better coincidence with those from the fit I
than those from the fit II. This probably results from the fact that
the constraint from the observation in the fit II is weak comparing
with the other two fits.

Furthermore, first impression on the impact of mass model
uncertainty can be made by comparing the different panels in
Fig.~\ref{fig3} for the fit I. In the region $38 \leqslant Z
\leqslant 82$, the simulations using the DZ10, DZ28, and KTUY mass
models better reproduce observations and the rms values with respect
to observations are approximately 0.25. For simulations using other
mass models, their rms values are 0.47, 0.40, 0.36, 0.35, 0.34 for
the RMF, ETFSI-2, ETFSI-Q, FRDM, HFB-17 simulations, respectively.

In the region $38 \leqslant Z < 56$, all simulations using different
mass models well reproduce the observations. However, the calculated
abundances for elements around $Z\thicksim 50$ are very different. A
possible reason is that the exact locations of nuclear shape
transition differ in different theoretical predictions, resulting in
different $S_n$ predictions and thus r-process
paths~\cite{Sun2008PRC}. In the region $56 \leqslant Z < 76$, large
troughs around $Z\thicksim 70$ are found for the simulations using
the ETFSI-2, ETFSI-Q and RMF mass models, which might be due to the
large neutron shell gap.

In the region $76 \leqslant Z \leqslant 82$, large abundances of Hg
(Z=80) are obtained for the FRDM and RMF simulations. This can be
traced back to the longer irradiation times for components with
higher neutron density. Moveover, the Pb abundances for different
simulations also deviate each other and the lowest value is obtained
for the DZ10 simulation. In fact, the relative Pb abundance from CS
22892-052 is not consistent with that from CS31082-001, where the
former is about 6 times larger than the latter. Therefore, precise
stellar data of elements heavier than Au in the future can further
improve the calculations of abundances of actinides, and can be also
used to test the mass predictions for the corresponding heavy
elements.

\subsection{The Influence of $\beta$-decay Rates}

Besides the physical uncertainty in nuclear masses for $r$-process
calculations, another important source of physical uncertainty is
$\beta$-decay rates of neutron-rich nuclei. To investigate its
influence on age estimates, we have further performed $r$-process
calculations by replacing the FRDM+QRPA $\beta$-decay
predictions~\cite{FRDMQRPA} with those from the ETFSI+CQRPA
model~\cite{ETFSICQRPA} if available. The astrophysical conditions
are readjusted to reproduce the observations when the ETFSI+CQRPA
$\beta$-decay predictions are included.

\begin{figure}
  \includegraphics[width=12cm]{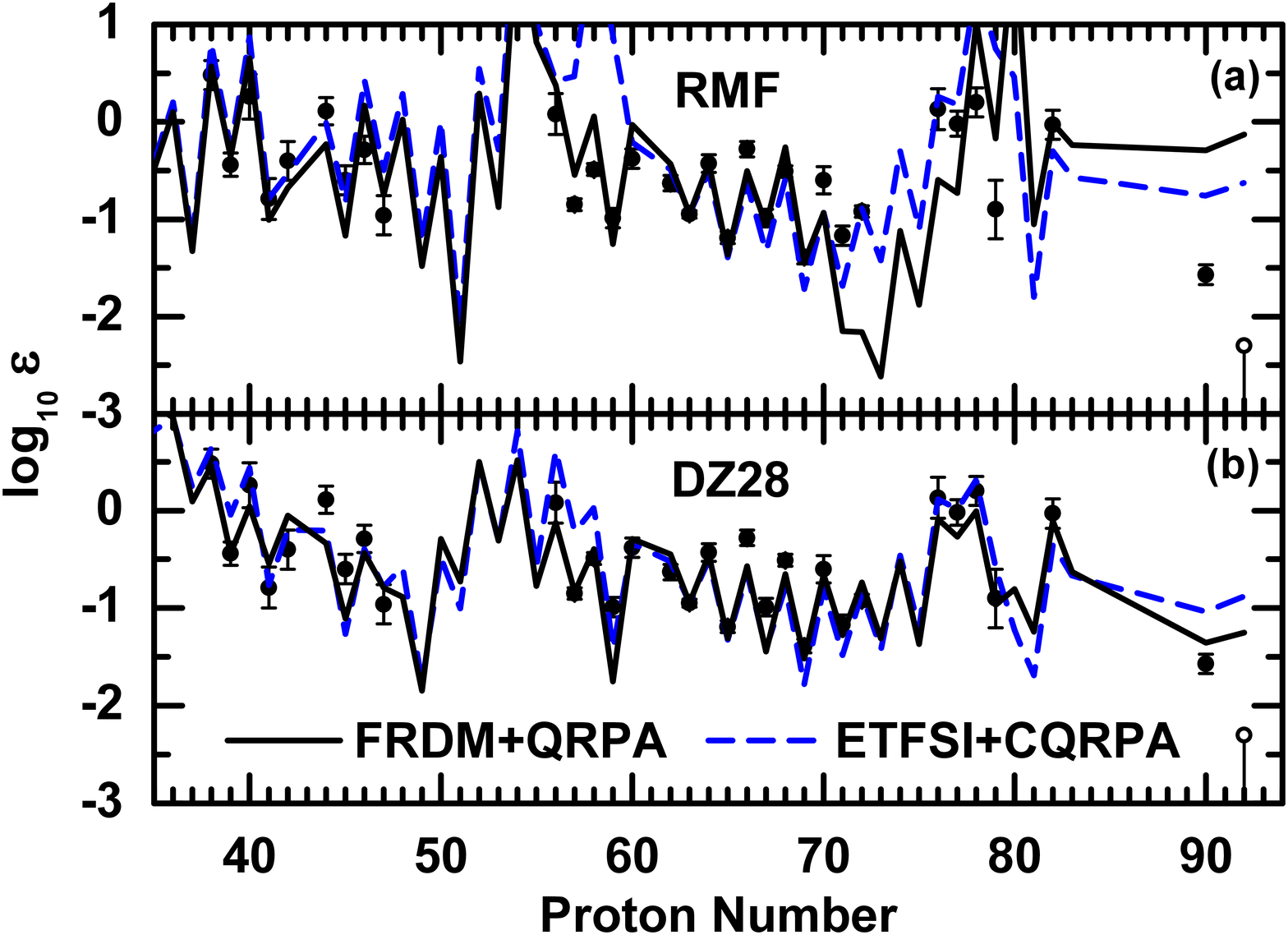}\\
  \caption{(Color online) Calculated $r$-process abundances using the $\beta$-decay rates
    predicted in the FRDM+QRPA (solid curve) and ETFSI+CQRPA (dashed curve) methods.
    The calculated abundances (scaled to Eu) for the RMF and DZ28 mass models are shown
    in the panel (a) and panel (b), respectively. The filled and open circles are the same
    as in Fig.~\ref{fig3}.}\label{fig4}
\end{figure}

Fig.~\ref{fig4} presents the corresponding results from the fit I,
where the solid and dashed curves are the predicted $r$-process
abundances using $\beta$-decay rates from the FRDM+QRPA and
ETFSI+CQRPA models, respectively. Two representative models, DZ28
and RMF, are used as examples and their corresponding simulations
are shown in the panel (a) and panel (b), respectively. As the data
from the ETFSI+CQRPA model are only available for nuclei with
neutron numbers around the magic numbers, the differences of the
predicted $r$-process abundances using these two $\beta$-decay rates
are observed mainly for elements with $Z \thicksim 56$ (with $N
\thicksim 82$) and $Z \thicksim 76$ (with $N \thicksim 126$). It is
found that the simulations using FRDM+QRPA can better reproduce the
observed abundances of elements with $Z \thicksim 56$, while the
abundances determined using ETFSI+CQRPA are generally in better
agreement with the observed abundances of elements with $Z \thicksim
76$.

The differences of $\log_{10}$(Th/U) due to the variance of the
$\beta$-decay rates can be as large as 0.25 for the FRDM simulation,
which corresponds to an age variation of about 5.5 Gyr, which
clearly shows the important influence of the $\beta$-decay rates on
the Th/U chronometer.

\subsection{Age Determination from Th/U Chronometer}

\begin{figure}
  \includegraphics[width=12cm]{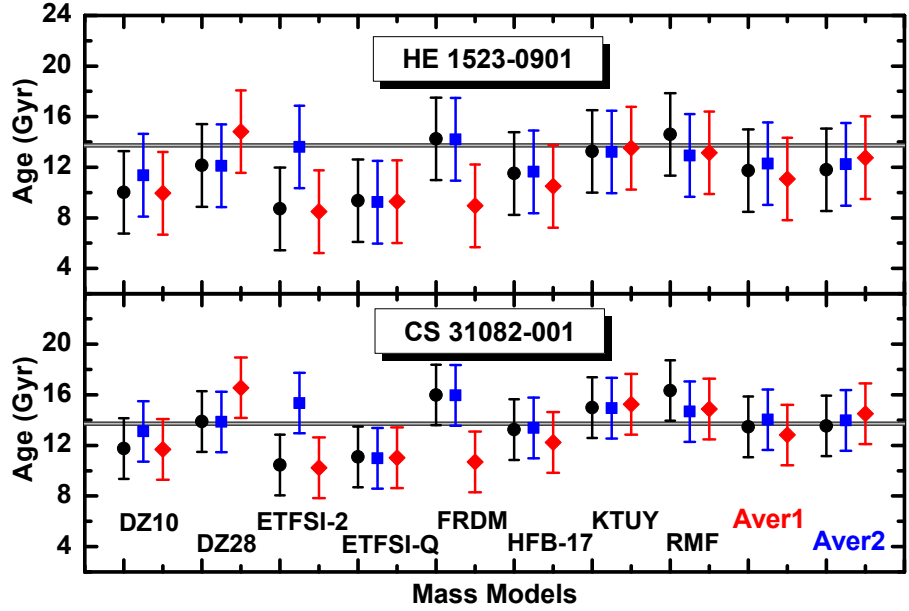}\\
  \caption{(Color online) The ages of the metal-poor stars HE 1523-0901
    and CS 31082-001 determined using various mass
    models and fits. The circles (squares, diamonds)
    denote ages determined using the fit I (II, III).
    The average ages with all mass models and the three mass
    models DZ10, DZ28, KTUY are labeled as Aver1 and Aver2,
    respectively. The shadowed area
    corresponds to the age of the universe~\cite{WMAP}.
    }\label{fig5}
\end{figure}

With the initial $r$-process abundance ratio Th/U and their present
observed values, one can eventually deduce the age of low
metallicity stars. In Fig.~\ref{fig5}, the ages of the metal-poor
stars HE 1523-0901 and CS 31082-001 are summarized. The circles
(squares, diamonds) denote ages determined using the fit I (II,
III). The average ages for all mass models and the three mass models
DZ10, DZ28, HFB-17 are labeled as Aver1 and Aver2, respectively. The
errors are attributed to uncertainties from observations of the
corresponding metal-poor stars. The shadowed area corresponds to the
age of the universe derived from WMAP data~\cite{WMAP}.

In general, for a given mass model, the ages determined using
different fits are in good agreements with each other. The
discrepancy is generally within 2.7 Gyr. This reflects that the age
estimate is not sensitive to different fits, which correspond to the
possible astrophysical conditions of $r$-process calculation.
Similar conclusion is also obtained based on the dynamical
$r$-process model in Ref.~\cite{Otsuki2003NA}. As for the ETFSI-2 or
FRDM models, the production ratio Th/U predicted from the fit III is
too large to give a reasonable age for the solar system (refer to
Fig. \ref{fig2}).

On the other hand, for a fixed fit, the deviation using different
mass models is rather large. The corresponding difference can be as
large as 5.9, 5.0, and 6.3 Gyr for the fit I, II, and III,
respectively. From the above discussion, it is clear that the DZ10,
DZ28, and KTUY simulations better reproduce the stable element
abundances than the others. Furthermore, their predictions for Th/U
production ratio are also consistent with the constraint from the
Th/U ratio measured in the solar system. Therefore, these mass
models might be more credible and are selected to estimate the age
of the metal-poor star.

In order to estimate the age of the star more reliably, one could
adopt the average values for different simulations. The
corresponding uncertainty is their root mean square deviation. In
this way, the discrepancy between the average values of ages
estimated using the three selected mass models for different fits
decreases to 0.9 Gyr. For the fit I, the age of the metal-poor star
HE 1523-0901 is obtained as $11.8 \pm 3.7$ Gyr which agrees with the
value of $13.2$ Gyr in Ref.~\cite{HE15230901}. The uncertainty of
age determined here includes that of 1.6 Gyr from nuclear mass
models and 3.3 Gyr from observation. Similarly, ages of CS 31082-001
and BD +17$^\circ$3248 are deduced as $13.5 \pm 2.9$ Gyr and $10.9
\pm 2.9$ Gyr, respectively. These results can served as the lower
limit of the cosmic age, and are in consistent with results derived
from the WAMP data~\cite{WMAP}.

For comparison, the corresponding average ages for all the mass
models using the fit I are determined as $11.7 \pm 4.0$, $13.5 \pm
3.3$, and $10.9 \pm 3.3$ Gyr for HE 1523-0901, CS 31082-001, and BD
+17$^\circ$3248, respectively, with uncertainties of 2.2 Gyr from
nuclear mass models. In Table~\ref{tbl}, we have tabulated the
predicted abundances of Th and U, the production ratios Th/U, and
the ages of the metal-poor stars HE 1523-0901, CS 31082-001, BD
+17$^\circ$3248 for different fits.

\section{Summary and perspective}

In this paper, we have analyzed the $r$-process actinide production
and determined the ages of the metal-poor stars HE 1523-0901, CS
31082-001, BD +17$^\circ$3248. The predictions are made within the
framework of the classical $r$-process model. This approach is
especially suitable for evaluating uncertainties emerging from the
nuclear physics inputs and the astrophysical conditions. In the
calculations, new developments in mass models and the updated
stellar data have been considered. By directly simulating the
observed abundances in the low-metallicity stars or the solar
system, the impact of the potential uncertainties in $r$-process
nucleosynthesis condition on the Th/U chronometer is investigated.
It is found that the age estimate is more sensitive to the mass
models than the fits. By adopting the three selected mass models
(i.e., DZ10, DZ28, KTUY), the ages of the metal-poor stars HE
1523-0901, CS 31082-001, and BD +17$^\circ$3248 are determined as
11.8, 13.5, and 10.9 Gyr, respectively, with an uncertainty of about
1.6 Gyr from nuclear mass models. Moreover, the impact of the mass
predictions and $\beta$-decay rates on the Th/U chronometer have
been discussed in details.

In order to distinguish the difference between various fitting cases
(thus possible different origins of elements in very metal-poor
stars), the stellar data, especially those missing abundances in the
medium heavy elements, are urgently required in our investigations.
Meanwhile, the stellar data of heavy elements heavier than Au, also
play a very important role in $r$-process chronometers, and can be
used as a test for the corresponding mass predictions of heavy
isotopes in different mass models.

\section{Acknowledgements}

We would like to thank F. Montes for valuable discussions. This work
is partly supported by Major State 973 Program 2007CB815000 and the
NSFC under Grant Nos. 10775004, 10975008 and 10947149.


\begin{table}
\begin{center}
\caption{The initial abundances of $^{232}$Th and $^{238}$U, the
production ratios $^{232}$Th$/^{238}$U, and the ages of the
metal-poor stars HE 1523-0901, CS 31082-001, BD +17$^\circ$3248 (in
Gyr) calculated using different fits and mass models. The
$\beta$-decay rates are taken from FRDM+QRPA if the experimental
values are not available.}\label{tbl}
\begin{tabular}{c|c|cccccccc}
\hline \hline
Fit          &Mass Model  &~   &$^{232}$Th   &~~~    &$^{238}$U   &$^{232}$Th$/$$^{238}$U   &HE 1523-0901  &CS 31082-001    &BD +17$^\circ$3248\\
\hline
\multirow{8}*{I}         &DZ10        &~   &0.0643       &~~~    &0.0256      &2.51                     &10.01         &11.76           & 9.14\\
            &DZ28        &~   &0.1468       &~~~    &0.0732      &2.01                     &12.14         &13.88           &11.27\\
            &ETFSI-2     &~   &0.1125       &~~~    &0.0390      &2.88                     & 8.71         &10.45           & 7.84\\
            &ETFSI-Q     &~   &0.1007       &~~~    &0.0374      &2.69                     & 9.36         &11.10           & 8.48\\
            &FRDM        &~   &0.2009       &~~~    &0.1252      &1.60                     &14.25         &15.99           &13.38\\
            &HFB-17      &~   &0.1413       &~~~    &0.0659      &2.14                     &11.51         &13.25           &10.64\\
            &KTUY        &~   &0.2387       &~~~    &0.1338      &1.78                     &13.25         &14.99           &12.38\\
            &RMF         &~   &0.6605       &~~~    &0.4273      &1.55                     &14.60         &16.34           &13.73\\
\hline \hline
\multirow{8}*{II}        &DZ10        &~   &0.1153       &~~~    &0.0530      &2.18                     &11.37         &13.11           &10.50\\
            &DZ28        &~   &0.2023       &~~~    &0.1006      &2.01                     &12.11         &13.86           &11.24\\
            &ETFSI-2     &~   &0.0140       &~~~    &0.0081      &1.72                     &13.61         &15.35           &12.74\\
            &ETFSI-Q     &~   &0.6341       &~~~    &0.2326      &2.73                     & 9.24         &10.98           & 8.37\\
            &FRDM        &~   &0.0644       &~~~    &0.0400      &1.61                     &14.21         &15.95           &13.34\\
            &HFB-17      &~   &0.5475       &~~~    &0.2588      &2.12                     &11.64         &13.38           &10.76\\
            &KTUY        &~   &0.3495       &~~~    &0.1951      &1.79                     &13.21         &14.95           &12.34\\
            &RMF         &~   &0.0128       &~~~    &0.0069      &1.84                     &12.93         &14.67           &12.06\\
\hline \hline
\multirow{8}*{III}       &DZ10        &~   &0.0794       &~~~    &0.0314      &2.53                     & 9.94         &11.69           & 9.07\\
            &DZ28        &~   &0.1119       &~~~    &0.0741      &1.51                     &14.82         &16.56           &13.95\\
            &ETFSI-2     &~   &0.0866       &~~~    &0.0294      &2.95                     & 8.49         &10.23           & 7.62\\
            &ETFSI-Q     &~   &0.0464       &~~~    &0.0171      &2.71                     & 9.28         &11.02           & 8.41\\
            &FRDM        &~   &0.1228       &~~~    &0.0437      &2.81                     & 8.95         &10.69           & 8.08\\
            &HFB-17      &~   &0.1385       &~~~    &0.0580      &2.39                     &10.49         &12.23           & 9.62\\
            &KTUY        &~   &0.1331       &~~~    &0.0767      &1.74                     &13.51         &15.25           &12.64\\
            &RMF         &~   &0.1795       &~~~    &0.0995      &1.80                     &13.14         &14.88           &12.27\\
\hline \hline
\end{tabular}
\end{center}
\end{table}


\begin{thebibliography}{99}
\bibitem{footnote}Due to the significant short half-life of $^{235}$U
($\tau_{1/2}=7.04 \times 10^8$ yr) compared to $^{238}$U
($\tau_{1/2}=4.468 \times 10^9$ yr), we approximate the uranium
abundances that of $^{238}$U. For briefly, we'll use Th and U
represent $^{232}$Th and $^{238}$U hereafter.
\end{thebibliography}
\end{document}